\input harvmac
\input epsf
\def\p{\partial}
\def\ap{\alpha'}
\def\half{{1\over 2}}

\def\te{\tilde{\eta}}


\Title{}{\vbox{\centerline{Power Spectra in Spacetime Noncommutative Inflation}}}

\centerline{Qing-Guo Huang and Miao Li }
\centerline{\it Institute of Theoretical Physics}
\centerline{\it Academia Sinica, P. O. Box 2735}
\centerline{\it Beijing 100080}
\medskip
\centerline{\tt huangqg@itp.ac.cn}
\centerline{\tt mli@itp.ac.cn}

\bigskip
The spacetime uncertainty relation, which deviates from general
relativity, emerges in String/M theory. It is possible to observe
this deviation through cosmological experiments, in particular
through the measurements of CMB power spectrum. This paper extends
some previous observations to more general inflation schemes. We
find that the noncommutative spacetime effects always suppress the
power spectrum of both the scalar and tensor perturbations, and
may provide a large enough running of the spectral index to fit
the results of WMAP in the inflation model.

\Date{November, 2003}

\nref\wmap{H. V. Peiris et AL., astro-ph/0302225;
C. L. Bennett et AL., Astro-ph/0302207;
D. N. Spergel et al., astro-ph/0302209. }

\nref\low{P. Mukherjee, Y. Wang, astro-ph/0303211;
S. L. Bridle, A. M. Lewis, J. Weller and G. Efstathiou, astro-ph/0302306;
G. Efstathiou, astro-ph/0306431;
E. Gaztanaga, J. Wagg, T. Multamaki, A. Montana and D. H. Hughes,
astro-ph/0304178;
M. Kesden, M. Kamionkowski and A. Cooray, astro-ph/0306597. }

\nref\hlf{Q. G. Huang, Miao Li, JHEP 0306 (2003) 014, hep-th/0304203. }

\nref\tmb{S. Tsujikawa, R. Maartens, R. Brandenberger, astro-ph/0308169. }

\nref\hle{Q. G. Huang, Miao Li, astro-ph/0308458, JCAP 0311 (2003) 001. }

\nref\perg{A. Riotto, hep-ph/0210162;
V. F. Mukhanov, H. A. Feldman and R. H. Brandenberger, Phys. Rep. 215
(1992) 203;
E. D. Stewart and D. H. Lyth, Phys. Lett. B 302 (1993) 171-175;
L. Wang, V. F. Mukhanov and P. J. Steinhardt, Phys. Lett. B 414 (1997) 18-27. }

\nref\dis{
R. H. Brandenberger, J. Martin, Mod. Phys. Lett. A 16 (2001) 999-1006;
J. Martin, R. H. Brandenberger, Phys. Rev. D 63 (2001) 123501;
J. Martin, R. H. Brandenberger, Phys. Rev. D 65 (2002) 103514;
E. Keski-Vakkuri, M. S. Sloth, JCAP 0308 (2003) 001;
S. F. Hassan, M. S. Sloth, Nucl. Phys. B 674 (2003) 434-458, hep-th/0204110;
P. M. Ho, hep-th/0308103;
Giovanni Amelino-Camelia, Int. J. Mod. Phys. D 11 (2002) 35-60, gr-qc/0012051.}

\nref\ini{
U. H. Danielsson, JHEP 0207 (2002) 040, hep-th/0205227;
K. Goldstein and D. A. Lowe, Phys. Rev. D 67 (2002) 063502,
hep-th/0208167;
L. Bergstrom and U. H. Danielsson, JHEP 0212 (2002) 038,
hep-th/0211006;
R. Easther, B. R. Greene, W. H. Kinney and G. Shiu,
Phys. Rev. D 66 (2002) 023518, hep-th/0204129;
J. C. Niemeyer, R. Parentani and D. Campo,
Phys. Rev. D 66 (2002) 083510, hep-th/0206149;
V. Bozza, M. Giovannini and G. Veneziano, JCAP 0305 (2003) 001, hep-th/0302184;
C. Armendariz-Picon and E. A. Lim, hep-th/0303103;
D. J. H. Chung, A. Notari and A. Riotto, hep-ph/0305074;
N. Kaloper, M. Kleban, A. Lawrence, S. Shenker and L. Susskind,
JHEP 0211 (2002) 037, hep-th/0209231;
J. Martin, R. Brandenberger, hep-th/0305161;
C. P. Burgess, J. M. Cline, F. Lemieux and R. Holman, astro-ph/0306236;
O. Elgaroy and S. Hannestad, astro-ph/0307011;
G. L. Alberghi, R. Casadio and A. Tronconi, gr-qc/0303035.}

\nref\ncs{G. Veneziano, Europhys. Lett. 2, 199 (1986);
D. J. Gross and P. F. Mende, Nucl. Phys. B 303, 407 (1988);
D. Amati, M. Ciafaloni and G. Veneziano, Phys. Lett. B 216, 41 (1989);
R. Guida, K. Konishi and P. Provero, Mod. Phys. Lett. A 6, 1487 (1991). }
\nref\ncst{T. Yoneya, in " Wandering in the Fields ", eds. K. Kawarabayashi,
A. Ukawa ( World Scientific, 1987), p. 419;
M. Li and T. Yoneya, Phys. Rev. Lett. 78 (1977) 1219, hep-th/9611072;
T. Yoneya, Prog. Theor. Phys. 103, 1081 (2000), hep-th/0004074;
J. Polchinski, " String Theory " volume 2.}

\nref\cgg{C-S. Chu, B. R. Greene and G. Shiu, hep-th/0011241;
F. Lizzi, G. Mangano, G. Miele and M. Peloso, JHEP 0206 (2002) 049,
hep-th/0203099. }

\nref\bh{R. Brandenberger, P. M. Ho, Phys. Rev. D 66(2002) 023517,
hep-th/0203119. }

\nref\shiu{D. J. H. Chung, G. Shiu, M. Trodden, astro-ph/0305193. }

\nref\sal{S. M. Leach and A. Liddle, astro-ph/0306305. }

\nref\lk{R. Kallosh, A. Linde, hep-th/0306058; A. Linde, A. Riotto,
Phys. Rev. D 56 (1997) 1841-1844, hep-ph/9703209. }

\nref\mmj{M. Kawasaki, M. Yamaguchi, J. Yokoyama,
Phys. Rev. D 68 (2003) 023508, hep-ph/0304161;
M. Yamaguchi, J. Yokoyama, hep-ph/0307373;
B. Wang, Chi-Yong Lin, E. Abdalla, hep-th/0309175.}

\nref\run{B. Feng, M. Li, R. J. Zhang and X. Zhang, astro-ph/0302479;
J. E. Lidsey and R. Tavakol, astro-ph/0304113;
L. Pogosian, S. -H. Henry Tye, I. Wasserman and M. Wyman,
Phys. Rev. D 68 (2003) 023506, hep-th/0304188;
E. D. Grezia, G. Esposito, A. Funel, G. Mangano and G. Miele, gr-qc/0305050;
K. -I, Izawa, hep-ph/0305286;
S. Cremonini, hep-th/0305244;
E. Keski-Vakkuri and M. S. Sloth, hep-th/0306070;
S. A. Pavluchenko, astro-ph/0309834;
G. Dvali and S. Kachru, hep-ph/0310244. }

\nref\lowt{B. Feng and X. Zhang, astro-ph/0305020;
M. Fukuma, Y. Kono and A. Miwa, hep-th/0307029;
G. Efstathiou, astro-ph/0303127;
C. R. Contaldi, M. Peloso, L. Kofman and A. Linde, astro-ph/0303636;
J. M. Cline, R. Crotty and J. Lesgourgues, astro-ph/0304558;
S. DeDeo, R. R. Caldwell and P. J. Steinhardt, Phys. Rev. D 67 (2003) 103509;
A. D. Oliveira-Costa, M. Tegmark, M. Zaldarriaga and A. Hamilton,
astro-ph/0307282;
A. Lasenby and C. Doran, astro-ph/0307311;
X. J. Bi, B. Feng and X. Zhang, hep-ph/0309195;
J. Martin and C. Ringeval, astro-ph/0310382;
Yun-Song Piao, B. Feng and X. Zhang, hep-th/0310206. }

\nref\dhl{O. Dore, G. P. Holder and A. Loeb, astro-ph/0309281. }

Einstein's general relativity is broken by the quantum effects at
very short distances. If inflation happened before the hot big
bang, it may be possible to observe these quantum effects through
cosmological experiments, such as WMAP and SDSS, since the short
distances are stretched to the cosmic ones by the accelerating
expansion during inflation. If this is the case, it appears that
we need to well understand quantum gravity before we can calculate
the power spectrum of cosmological fluctuations accurately. Since
String/M theory is the most attractive framework of quantum
gravity, it is imperative to investigate the modification of the
primordial fluctuation spectrum in string theory. However, there
is no fundamental formulation of string theory in a time-dependent
background yet. We need resort to some general intuitions gained
from studying string theory. Spacetime uncertainty is one of such
intuitions and we shall in this paper extend some previous
observations concerning the effects caused by spacetime
uncertainty in the CMB power spectrum to a more general situation.

The first year results of WMAP \wmap\ put forward more restrictive
constraints on cosmological models and confirm the emerging
standard model of cosmology, a flat $\Lambda$-dominated universe
seeded by a nearly scale-invariant adiabatic Gaussian
fluctuations. WMAP also brings about some new intriguing results,
such as a running spectral index of the scalar fluctuations and an
anomalously low quadrupole of CMB angular power spectrum \low.
According to \refs{\hlf-\hle}, spacetime noncommutative power-law
inflation naturally produces a large enough running of the
spectral index to match the results of WMAP. However the large
suppression of $C^{TT}_2$ remains mysterious.

It is generally accepted that the large scale structures of the
universe originated from some small seed perturbations \perg,
which over time grew to become all of the structures we observe
today. Quantum fluctuations of the inflaton field are excited
during inflation and stretched to cosmological scales with the
expansion of our universe. At the same time, being the inflaton
fluctuations, ripples in spacetime are also excited and stretched
to cosmological scales. The physical wavelengths of these
fluctuations are so short that they are sensitive to the physics
at the very short distance during the period of inflation. Many
authors have discussed how the spectra of fluctuations in
inflationary cosmology depend on trans-Planckian physics and
whether these trans-Planckian effects can be observed. In general,
there are two ways to investigate them. One is that we can take
the trans-Planckian physics into account by modifying the
dispersion relation \dis. The other is called the minimal
trans-Planckian physics \ini, in which we calculate the power
spectrum of the fluctuations in inflationary cosmology, starting
with initial conditions imposed on mode by mode when the physical
wavelength of the fluctuation equals some critical length $l_n$
corresponding to a new energy scale $M_n =  l^{-1}_n$. In this
model, a mode of fluctuation is created when its wavelength became
equal to $l_n$, while the evolution equations of fluctuation modes
are unmodified. Thus the change is entirely encoded in the initial
conditions and there is no need to postulate some ad-hoc
trans-Planckian physics. As a result, the corrections can be
generally expanded in terms of some powers of $H / M_n$, where $H$
is the Hubble parameter during inflation. Thus we can probe  new
physics in this way only if the Hubble parameter $H$ during
inflation is not much smaller than the new physical scale $M_n$.

In perturbative string theory, the fundamental degree of freedom
is fundamental string with length scale $l_s$ as the minimal
physical length scale, implying a stringy uncertainty relation
\ncs: $\bigtriangleup x_p \bigtriangleup p \geq 1 + l^2_s
\bigtriangleup p^2$ or $\bigtriangleup x_p \geq l_s$.
Heuristically the minimal distance we can measure by using
fundamental string must not be shorter than its length $l_s$ since
the fundamental string is extensive. In nonperturbative string
theory or M theory, new degrees of freedom such as D-branes and
black holes must be taken into account, their effects are
suppressed by a factor such as $\exp(-1 / g_s)$ ($g_s \rightarrow
0$ in perturbative string theory), where $g_s$ is the string
coupling. A D-brane probe with a sufficiently small velocity can
be used to probe distances shorter than the string scale. In any
physical process a new uncertainty relation \ncst,
\eqn\ustc{\bigtriangleup t_p \bigtriangleup x_p \geq l^2_s, } is
observed, where $t_p$ and $x_p$ are the physical time and space.
We believe that M theory should not favor a ``hard" cut-off, but an
uncertainty relation between space and time. However it is
hard to incorporate this relation directly in string theory in a
fundamental formulation. On a general ground, if inflation is
affected by physics at a scale close to string scale or a related
scale, one expects that spacetime uncertainty must leave traces in
the CMB power spectrum \refs{\cgg, \bh} (see also
\refs{\hlf-\hle}). Indeed, an explicit calculation based on the
model proposed in \bh\ shows that the effects can be observed
\hlf.

Before we discuss the noncommutative inflation in detail, let us
sketch the physics behind this scenario. As in the usual inflation
model, we assume that the universe went through a slow-roll period
during which cosmic perturbations were generated. The larger the
scale of the perturbation, the earlier its generating time. The
long wave-length perturbations were generated earlier and crossed
out the horizon earlier, and re-entered the horizon later. For an
earlier generating time, the spacetime uncertainty relation tells
us that the uncertainty in creating time is smaller. Nevertheless,
the ratio of the time uncertainty and the creation time is larger,
so we expect that the correction of the power spectrum is larger
for a smaller wave-number. In addition, the usual power-law of the
spectrum tells us that a smaller wave-number comes with a smaller
power, and due the the larger correction of the relative
generating time, we expect that the noncommutative correction
makes the power of a smaller wave-number even smaller, thus
induces a larger spectral index. In the UV end, the story is
reversed, so the spectral index for a larger wave-number is
smaller, smaller than 1 when the wave-number is larger than a
``critical value".

The spacetime noncommutative effects can be encoded in a new
product, start product, replacing the usual algebra product. The
evolution of a homogeneous background will not change and the
standard cosmological equation of the inflation based on
Friedmann-Robertson-Walker (FRW) metric remains the same: \eqn\mp{
\ddot \phi + 3 H \dot \phi + V'(\phi) = 0, } \eqn\hu{ H^2 =
\left({\dot a \over a} \right)^{2} = {1 \over 3 M_p^2} \left(
\half {\dot \phi}^2 + V(\phi) \right), } here $M_p$ is the reduced
Planck mass and we assume the universe be spatially flat and the
inflaton $\phi$ be spatially homogeneous. If $\dot \phi^2 \ll
V(\phi)$ and $\ddot \phi \ll 3 H \dot \phi$, the scalar field
shall slowly roll down its potential. Define some
slow-roll parameters, \eqn\ep{\epsilon= - {\dot H \over H^2} =
{M_p^2 \over 2} \left({V' \over V}\right)^2, } \eqn\et{\eta =
\epsilon - {\ddot H \over 2 H \dot H} =M_p^2 {V'' \over V}, }
\eqn\xx{\xi^2=7 \epsilon \eta- 5 \epsilon^2-2 \eta^2+ \zeta^2
=M_p^4 {V' V''' \over V^2}, } where $\zeta^2={(d^3 H / dt^3) / (2
H^2 \dot H)}$, then the slow-roll condition can be expressed as
$\epsilon, \eta \ll 1$.

As impressive successes, inflation model not only easily solves
the flatness problem, the entropy problem and the horizon problem
in hot big bang model, but also provides a reasonable primordial
cosmological fluctuations. In order for structure formation to
occur via gravitational instability, there must have been small
preexisting fluctuations on physical length scales when they
crossed inside the Hubble radius in the radiation-dominated and
matter-dominated eras. We will consider small perturbations away
from the homogeneous and isotropic reference spacetime, the
spatially flat Friedmann-Robertson-Walker (FRW) spacetime in our
case. These metric perturbations can be decomposed into different
spin modes. The key issue
is that general relativity is a gauge theory where the gauge
transformations are general coordinate transformations from a
local reference frame to another, thus we need to define some
gauge invariant quantities, such as the tensor fluctuation
$h_{ij}$ and the comoving curvature perturbation $\cal R$, to
describe the cosmological density perturbations.

We will focus on the scalar perturbation in the spacetime
noncommutative inflation first. In the conformal coordinates, the
linear scalar perturbations of the metric can be expressed most
generally by two scalar degrees of freedom $A$ and $\psi$ and the
line-element becomes \eqn\pmt{ds^2 = a^2 [(1 + 2 A) d\chi^2 - (1 -
2 \psi) \delta_{ij} dx^i dx^j], } where the conformal time $\chi$
is defined as \eqn\ct{\chi =\int a^{-1} dt. } Since the stress
tensor does not have any non-diagonal component, we have $\psi =
A$. We define the comoving curvature perturbation as
\eqn\ccp{{\cal R} = \psi + H {\delta \phi \over \dot \phi}, }
which is gauge invariant, where $\delta \phi$ is the quantum
fluctuation of inflaton.

For convenience we introduce another time coordinate $\tau$ in the
noncommutative spacetime such that the metric becomes \eqn\met{d
s^2 = d t^2 - a^2(t) d {\vec x}^2 = a^{-2}(\tau) d \tau^2 -
a^2(\tau) d {\vec x}^2. } Now the uncertainty relation \ustc\
becomes \eqn\usr{ \bigtriangleup \tau \bigtriangleup x \geq l^2_s,
} where $x$ is the comoving spatial coordinate. Skipping the
detailed discussions in \bh, we simply write down the action of the perturbation which
incorporates the noncommutative case in four dimensions
\eqn\ap{S = V \int_{k < k_0} d \te d^3 k \half z^2_k(\te) (u'_{-k}
u'_{k} - k^2 u_{-k} u_{k}), }
where
\eqn\scaf{\eqalign{z_k^2(\te)&=z^2y_k^2(\te), \quad
y_k^2=(\beta_k^+\beta_k^-)^{\half},\cr {d\te \over
d\tau}&=\left({\beta_k^-\over \beta_k^+}\right)^\half,\quad
\beta_k^\pm =\half (a^{\pm 2}(\tau+\l_s^2k)+a^{\pm
2}(\tau-l_s^2k)),}} here $l_s$ is the string length scale,  $z = a
\dot \phi / H$, ${\cal R}_k = u_k(\te) / z_k(\te)$,
$k_0=(\beta_k^+ /\beta_k^-)^{1/4} l_s$ and the prime denotes
derivative with respect to the modified conformal time $\te$. Thus
the equation of the scalar perturbation can be written as
\eqn\eqp{u_k'' + \left(k^2 - {z''_k \over z_k}\right) u_k =0. }
After a lengthy but straightforward calculation, we get (up to the
first order of the slow-roll parameters and $\mu$ defined as below
to describe the spacetime noncommutative affects)
\eqn\zk{{z_k''
\over z_k } = 2 (a H)^2 \left(1 + {5 \over 2}\epsilon - {3 \over
2} \eta -2 \mu\right), } where $\mu = {H^2 k^2 / (a^2 M^4_s)}$,
$k$ is the comoving fourier mode and $M_s = l_s^{-1}$ is the
string mass scale. In this paper we only consider the case in
which the perturbations are generated inside the horizon. For the
slow-roll inflation, the conformal time $\chi$ can be
approximately integrated out from equation \ct, \eqn\cont{\chi =
\int {d t \over a} = \int {d a \over a^2 H} \approx {-1 \over a H}
(1 + \epsilon). } And from the third equation in \scaf, we get
\eqn\con{\chi \simeq (1+\mu)^{-1} \te, } therefore \eqn\ck{a H
\simeq {-1 \over \chi} (1 + \epsilon) \simeq {-1 \over \te} (1 +
\epsilon + \mu). } Using equation \zk\ and \ck, we obtain from
equation \eqp\ \eqn\zkk{u_k'' + \left(k^2 - {1 \over \te^2}
\left(\nu^2 -{1 \over 4}\right)\right) u_k =0, } where $\nu = {3
\over 2} + 3 \epsilon -\eta$. We notice that this equation is
similar to the commutative case and the only difference is that
the conformal time $\chi$ is replaced by the modified conformal
time $\te$. Next, we choose the initial conditions \eqn\ic{u_k =
{1 \over \sqrt{2 k}} e^{-i k \te}. }
The solution of equation
\zkk\ is
\eqn\sole{u_k=\half \sqrt{\pi}
e^{i(\nu+\half) \pi/2} (-\te)^{1/2} H_\nu^{(1)}(-k\te),} where
$H_\nu^{(1)}$ is the Hankel's function of the first kind. At the
superhorizon scales ($k^2 \ll {z_k'' \over z_k}$), the solution
can be expressed as \eqn\sol{u_k \simeq {1 \over \sqrt{2k}} (- k
\te)^{\half - \nu} \simeq {1 \over \sqrt{2k}} \left({k \over a
H}\right)^{\half - \nu} (1+\mu)^{\half - \nu}. }
After a simple
calculation, we have, from the second and forth equation of \scaf,
\eqn\yk{y_k \simeq 1 + {H^2 k^2 \over a^2 M_s^4} = 1 + \mu. } Thus
the power spectrum on superhorizon scales of the comoving
curvature can be expressed as \eqn\asc{P_s \simeq {k^3 \over 2
\pi^2} \left|{u_k \over z_k(\te)}\right|^2 \simeq {1 \over 2
\epsilon}{1 \over M_p^2} { \left({H \over 2 \pi} \right)^2}
\left({k \over aH}\right)^{2\eta-6\epsilon} (1+\mu)^{-4 - 6
\epsilon + 2 \eta}, } here $\te$ is the time when the fluctuation
mode $k$ crosses the Hubble radius ( $z_k'' / z_k = k^2$ ).
Plugging this condition into equation \zk, we get \eqn\ch{k^2 = 2
(a H)^2 \left(1+{5\over 2}\epsilon -{3 \over 2}\eta-2\mu \right).
} Or perturbatively up to the first of the slow-roll parameter
$\epsilon$, $\eta$ and the noncommutative parameter $\mu$, we get
\eqn\dk{d \ln k = (1-\epsilon+4 \epsilon \mu) H dt, } and
\eqn\dmu{ {d \mu \over d \ln k} = (1 + \epsilon - 4 \epsilon \mu)
{1 \over H} {d \over dt} \left({H^2 k^2 \over a^2 M^4_s}  \right)
\simeq - 4 \epsilon \mu. } Using equation \asc, \dk\ and \dmu, we
find the spectral index of the scalar perturbation and its running
\eqn\pe{n_s-1 \equiv s={d \ln P_s \over d \ln k } =-6 \epsilon+2
\eta + 16 \epsilon \mu, } \eqn\er{{d n_s \over d \ln k}=-24
\epsilon^2 + 16 \epsilon \eta -2 \xi^2 -32 \epsilon \eta \mu. }
When $l_s \rightarrow 0$ or $M_s \rightarrow +\infty$, the
noncommutative parameter $\mu = {H^2 k^2 / (a^2 M_s^4)}
\rightarrow 0$, equation \pe\ and \er\ reproduces the results in
commutative case.

Inflation predicts that there are also tensor perturbations
of the metric. Because the amplitude of primordial tensor
perturbations of the metric during the period of inflation only
depends on the energy scale of the inflation, we can determine the
energy scale of a model by the measurement of the amplitude of the
primordial gravitational wave perturbations. In
general the linear tensor perturbations may be written as
\eqn\gw{ds^2 = a^2(\chi) \left(d \chi^2 - (\delta_{ij} + 2 h_{ij})
dx^i dx^j \right), } where $|h_{ij}| \ll 1$. The tensor $h_{ij}$
has six degrees of freedom, but the tensor fluctuations are
traceless and transverse. With these four constraints, there
remain two physical degrees of freedom, or polarizations. Notice
that tensors $h_{ij}$ are gauge-invariant and therefore represent
two physical degrees of freedom. Here the stress-energy momentum
tensor is diagonal, as the one provided by the inflation
potential. Thus the tensor mode does not have a source in its
equation of motion and then, in the commutative case, the action
of the tensor mode is simply \eqn\nt{S = {M_p^2 \over 2} \int d^4
x \sqrt{-g} \p_\sigma h_{ij} \p^\sigma h_{ij}, } similar to the
action of two independent massless scalar fields. Using the same
argument as in the scalar perturbations in \bh, in the
noncommutative spacetime background, the gauge-invariant tensor
amplitude can be expressed as \eqn\vv{h_k = { v_k \over a_k M_p},
} where $a_k = a y_k$, where $y_k$ is given in equation \yk. The
motion equation of $v_k$ becomes \eqn\tv{v_k'' + \left( k^2 -
{(a_k)'' \over a_k } \right) v_k =0. }
After a lengthy
calculation, we find
\eqn\ay{{(a_k)'' \over a_k } = 2 (aH)^2
\left(1 - {\epsilon \over 2} -2 \mu \right). } and \eqn\tvv{v_k''
+ \left( k^2 - {1 \over \te^2} \left(\nu^2 - {1 \over 4} \right)
\right) v_k =0 ,}
where $\nu = 3/2 + \epsilon$. Similar to the
scalar fluctuation, the tensor modes at superhorizon scales can be
solved \eqn\st{v_k \simeq {1 \over \sqrt{2k}} \left({k \over aH}
\right)^{\half - \nu} (1 + \mu)^{\half - \nu}. }
In the end, the
amplitude of the metric tensor perturbations is found to be
\eqn\tenp{P_T = 2 \times {k^3 \over 2 \pi^2} |h_k|^2 = 2 \times
{k^3 \over 2 \pi^2} {1 \over M_p^2}\left| {v_k \over a_k}\right|^2
=2 \times {1 \over M_p^2} \left({H \over 2 \pi}\right)^2 \left( {k
\over aH}\right)^{-2 \epsilon} (1+\mu)^{-4-2\epsilon}, } where the
factor 2 comes from the number of the independent physical degrees
of freedom of the tensor perturbation $h_{ij}$. Using eqs. \asc\
and \tenp, we find \eqn\ts{P_T \simeq 4 \epsilon P_s. } In
power-law inflation $\epsilon = 1 / p$ where p is the power of the
time in the scale factor of the universe, thus $P_T = (4 / p)
P_s$, the same as the result in \tmb. The spectral index $n_T$ of
the tensor perturbations and its running $d n_T / d \ln k$ are
\eqn\sitp{n_T = {d \ln P_T \over d \ln k} = -2 \epsilon + 16
\epsilon \mu, } and \eqn\srtp{{d n_T \over d \ln k} = -8
\epsilon^2 + 4 \epsilon \eta - 32 \epsilon \eta \mu. } These two
equations also reproduce the results in commutative case when $\mu
\rightarrow 0$. The spectrum of the tensor perturbations in
commutative case must be red ($n_T < 1$), since the slow-roll
parameter $\epsilon$ is positive. In noncommutative spacetime, it
can be blue when $\mu > 1/8$. We find that the amplitude of the
tensor perturbations are also suppressed in noncommutative
spacetime.

From equation \asc\ and \tenp, we see that the spacetime
noncommutative effects suppress the power spectrum of the
primordial scalar and tensor perturbations approximately with a
same factor $(1+\mu)^{-4}$, leading to a more blue spectrum with a
correction $+16 \epsilon \mu$ appearing in \pe\ and \sitp, where
$\epsilon$ must be positive by definition in \ep. Using the
argument in \bh, we find that the only difference in the
noncommutative case from the commutative case is replacing the
conformal time $\chi$ with modified conformal time $\te$ and $\chi
\simeq (1 + \mu)^{-1} \te$ in equation \con. And $\te$ plays the
role of the conformal time in the commutative case. Thus we can
get the noncommutative result by replacing $\chi$ in the
commutative case by $(1 + \mu)^{-1} \chi$ which means a delay of
the time when the fluctuation mode cross outside the horizon,
since both of the conformal and the modified conformal time are
negative and $(1 + \mu)^{-1} <1$. After a simple calculation, we
find that the Hubble constant becomes smaller in the
noncommutative case than in the commutative case by a factor $(1 +
\mu)^{-1}$. Since the power spectrum $P \sim H^2 / (M_p y_k)^2$
and $y_k = 1 + \mu$ in eq.\yk, we predict that the power spectrum
in the noncommutative case is suppressed by a factor $(1 +
\mu)^{-4}$.

We also note that in a de Sitter space, the noncommutative
inflation is the same as the commutative one, since $\epsilon =
0$. The reason is that the Hubble constant is not varying with
time and there are no time delaying effects due to the
noncommutative effects.

Before we discuss how to match the results of the
WMAP group \wmap, we briefly review their results. For the scalar
modes, the spectral index and its running at two different scales
are
\eqn\map{\eqalign{ n_s&=0.93\pm
0.03, \quad {dn_s\over d\ln k}=-0.031^{+0.016}_{-0.017} \quad
\hbox{at}\quad k=0.05 \hbox{Mpc}^{-1},\cr n_s&=1.20^{+
0.12}_{-0.11}, \quad {dn_s\over d\ln k}=-0.077^{+0.050}_{-0.052}
\quad \hbox{at}\quad k=0.002 \hbox{Mpc}^{-1}}. } Here we use the
new results revised by Peiris et al. on 12, May. The data of
the WMAP give rise to a maximum of the tensor/scalar ratio leading
to a constraint on the slow-roll parameter $\epsilon < 1.28/16 =
0.08$ ($95 \% CL$).

If we want to directly fit the data of WMAP by using the inflation
model in noncommutative spacteime, we need to calculate the
primordial power spectrum exactly, use some sophisticated
techniques, such as running CMBfast, to compute the $C_{l}^{TT}$
etc and compare them with the data directly with taking the
likelihood into account. In our paper, we want to show that we can
produce the large enough running of the spectral index of the
power spectrum in the best-fit results of WMAP group.

We have showed that the spacetime noncommutative power-law
inflation can fit the spectral index and its running quite nicely
in \hlf\ and \hle. Here we extend our previous discussions to a
general slow-roll inflation model and discuss how to match the
index of the power spectrum and its running by using our previous
results. From equation \pe, we have $\eta = \half s + 3 \epsilon -
8 \epsilon \mu$. With this substituted into \er, it becomes
\eqn\ers{{d n_s \over d \ln k}=s^2+(13-48 \mu)s \epsilon +(28-304
\mu+ 512 \mu^2)\epsilon^2 -2 \zeta^2. } If spacetime is
commutative ($\mu = 0$), equation \ers\ becomes \eqn\aer{{d n_s
\over d \ln k}=s^2+13 s \epsilon+28 \epsilon^2-2 \zeta^2. } Thus
in commutative case, $\zeta^2$ must be very large in order to get
a large enough negative value of $d n_s / d \ln k$ in \aer,
specially when the CMB power spectrum is blue $(s > 0)$, to fit
the WMAP data, since $\epsilon >0$. However it is quite hard to
get a large $\zeta^2$ in the known typical inflation model \shiu\
and has been also discussed in \sal. If we take the noncommutative
effects into account, the second or the third term of equation
\ers\ will become negative for some suitable values of $\mu$,
there is no longer the need of a large $\zeta^2$. In practice, we
shall ignore $\zeta^2$ and show the constraint on values of
$\epsilon$ and of $\mu$ in order to fit the experimental data, in
fig. 1.

\bigskip
{\vbox{{\epsfxsize=10cm
        \nobreak
    \centerline{\epsfbox{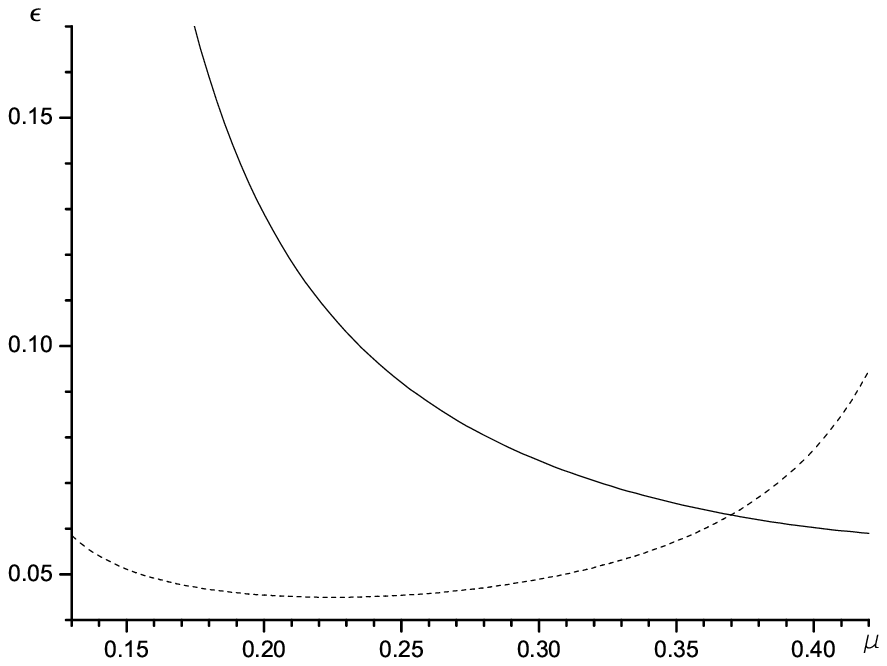}}
        \nobreak\bigskip
    {\raggedright\it \vbox{
{\bf Figure 1.}
{\it The value of the solid line is $n_s=1.20$, ${d n_s / d \ln k}=-0.077$
and the dash line is $n_s=0.93$, ${d n_s / d \ln k}=-0.031$ by using
equs.\hbox{\pe\ and \ers}, here we neglect $\zeta^2$.}
 }}}}
    \bigskip}
Further, the constraints on the parameters $\epsilon$ and $\mu$
are loosened after we take the contribution of $\zeta^2$ into
consideration. We see from fig.1 that it is possible to naturally
realize a large enough running of spectral index of the scalar
metric perturbations after we take the spacetime noncommutative
effects into consideration.

In the following we shall check some typical inflation models.
According to equation \pe\ \er\ \sitp\ and \srtp, we see that the
noncommutative term $\mu$ always appears in a product with the
slow-roll parameter $\epsilon$. Thus, there is a  significant
effect in the models only when $\epsilon$ is large. We will see
that the noncommutative effects in cases of the power-law model
and the chaotic model are large enough to realize the running of
the index of the power spectrum.

\noindent Case 1.

The potential $V=\lambda^4 \exp (-\sqrt{2/p} (\phi/M_p))$
leads to power-law inflation with $a \sim t^p$ and the slow-roll
parameters are $\epsilon = 1 / p $, $\eta = 2 / p $ and
$\xi^2 = 4 / p^2 $.
Since $\epsilon$ is a constant, we can integrate equation \dmu\ to get
\eqn\inte{{d \mu \over d \ln k} = -4 \epsilon \mu = - {4 \over p} \mu, }
and
\eqn\inm{\mu = \left({k \over k_c}\right)^{ -4 \epsilon}
= \left({k \over k_c}\right)^{-4 / p}. }
Equations \pe\ and \er\ become
\eqn\pep{n_s = 1 - {2 \over p} + {16 \over p} \mu =
- {2  \over p} + {16 \over p} \left({k \over k_c}\right)^{-4/p}, }
\eqn\erp{{dn_s \over d \ln k} = - {64 \over p^2} \mu
=- {64 \over p^2} \left({k \over k_c}\right)^{-4/p}. }
These formulas are exactly the same as in \hlf.

\noindent Case 2.

Potential $V=\lambda^4 (\phi/M)^p$ ($p \geq 2$). The number of
e-folds is $N \simeq {1 \over M^2_p} \int^{\phi_N}_{\phi_{end}} {V
\over V'} d \phi \simeq \phi_N^2 / (2 p M^2_p)$, so $\phi_N =
\sqrt{2 p N} M_p$. The slow-roll parameters are $\epsilon=(p^2/2)
M^2_p / \phi^2= p/(4N)$, $\eta = (p-1)/(2 N)$ and $\xi^2=
(p-1)(p-2)/(4 N^2)$. Thus the spectral index and its running
become \eqn\lamns{n_s-1 = s = {1 \over N } \left(-1 - \half (1-8
\mu) p \right), } and \eqn\lamnsk{{d n_s \over d \ln k} = - {1
\over N^2} \left(1 + {p \over 2} + 4 p (p-1) \mu \right). } For
$N=50$ and $p=2$, $n_s = 0.96$ and $ d n_s / d \ln k = -0.0008$,
the running is too small to fit the experimental data in the
commutative case ($\mu = 0$). Eq. \lamns\ also tells us that the
power spectrum must be red in this case. The situation changes in
the noncommutative spacetime background. When $\mu
> (1 + 2/p) / 8$, the spectrum becomes blue. For a small
$k$, the larger $\mu$ the earlier it crosses outside the horizon
and thus produces a larger spectral index. First, we try to fit
the data at $k = 0.05 Mpc^{-1}$ in eq. \map. We show the fitting
parameters in fig. 2 - 4.
\bigskip
{\vbox{{\epsfxsize=8cm
        \nobreak
    \centerline{\epsfbox{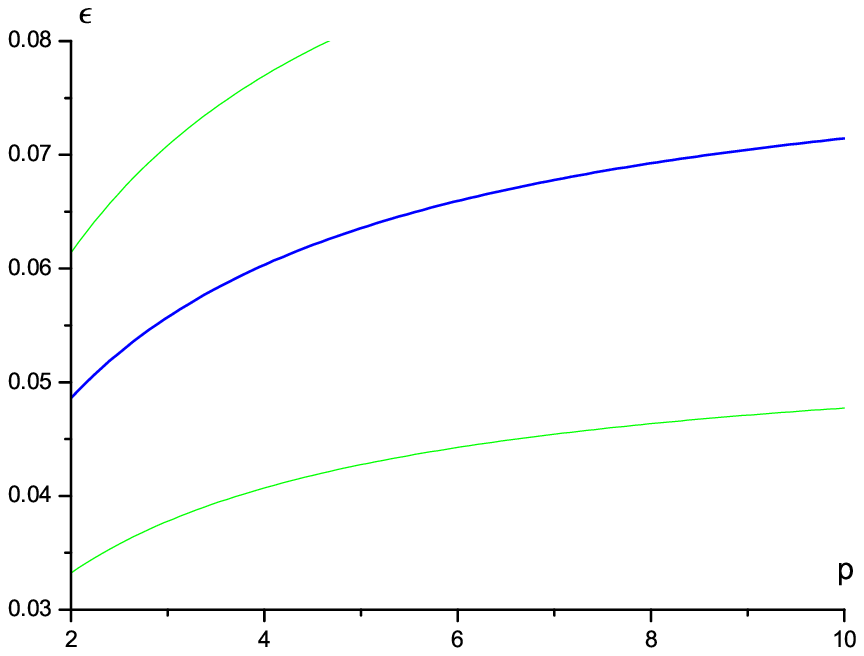}}
        \nobreak\bigskip
    {\raggedright\it \vbox{
{\bf Figure 2.}
{\it where $\epsilon$ is the slow-roll parameter with the constraints
$\epsilon < 0.08$ from WMAP and p is the index of the inflaton in the
potential.}
 }}}}
    \bigskip}

\bigskip
{\vbox{{\epsfxsize=8cm
        \nobreak
    \centerline{\epsfbox{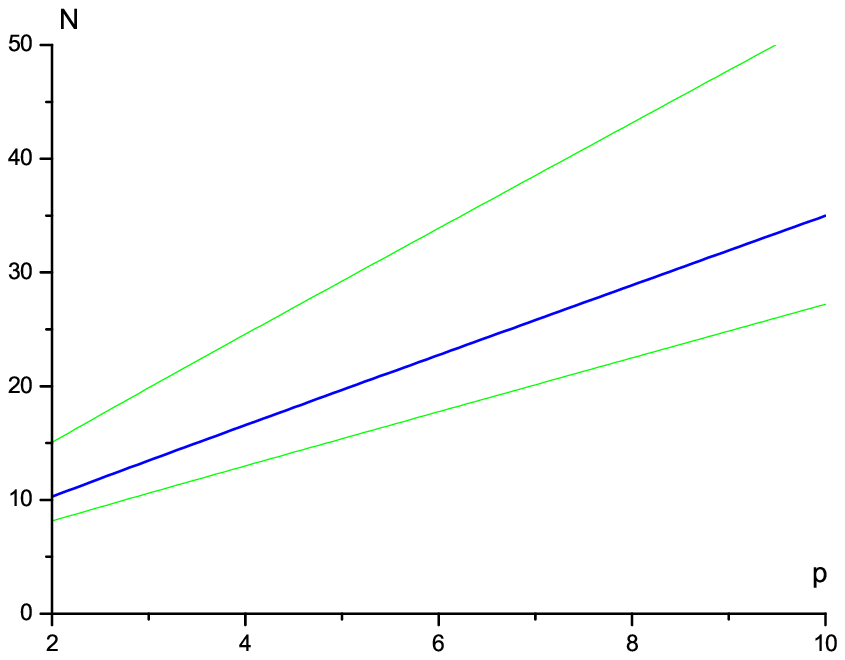}}
        \nobreak\bigskip
    {\raggedright\it \vbox{
{\bf Figure 3.}
{\it where $N$ is the e-folding number and p is the index of the inflaton
in the potential. }
 }}}}
    \bigskip}

\bigskip
{\vbox{{\epsfxsize=8cm
        \nobreak
    \centerline{\epsfbox{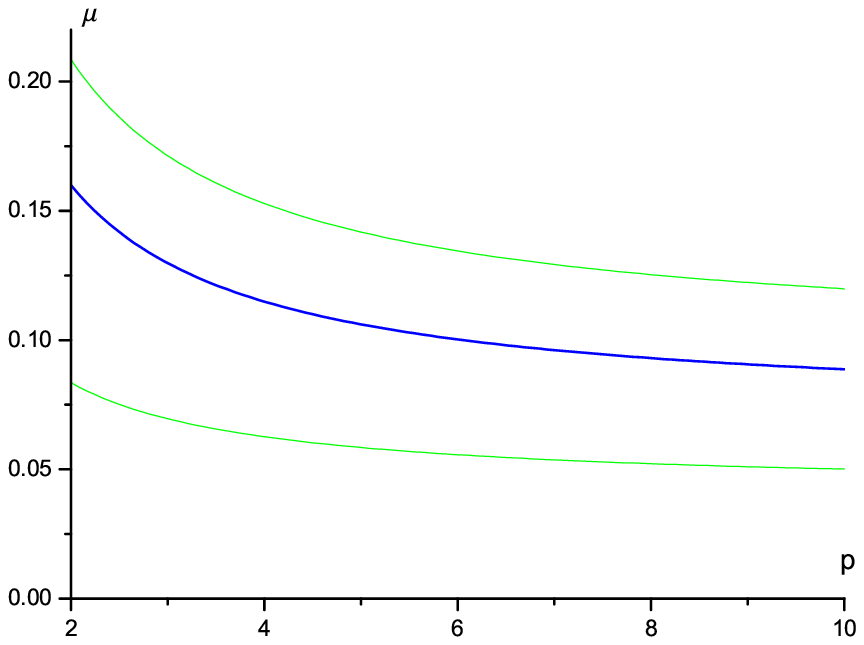}}
        \nobreak\bigskip
    {\raggedright\it \vbox{
{\bf Figure 4.} {\it where $\mu = {H^2 k^2 / (a^2 M_s^4)}$ is the
noncommutative parameter and p is the index of the inflaton in the
potential. }
 }}}}
    \bigskip}
The blue lines in fig. 2 - 4 correspond to $n_s = 0.93$ and $d n_s
/ d \ln k = -0.031$ and the green lines show the range of the
parameters to fit the likelihood of $n_s$ and $d n_s / d \ln k$.
In general, the larger the slow-roll parameter $\epsilon$, the
larger the noncommutative effects. From fig. 2-4, we choose $p =
10$ as an example in order to get a large e-folds number.
According to the blue lines in these three figures, we can get $N
= 35$, $\epsilon = 0.071$ and $\mu = 0.089$. Since $\mu = 0.089
\ll 1$, we can trust our perturbative results. We can also fit the
data at the mode $k = 0.002 Mpc^{-1}$ and the results are $N =
42$, $\epsilon = 0.060$ and $\mu = 0.36$. We can see that the
spectrum runs from the blue one ($n_s = 1.20$) at $N = 42$ to the
red one ($n_s = 0.93$) at $N = 35$. This example shows us that the
chaotic inflation model can fit the results of WMAP in
noncommutative spacetime.

\noindent Case 3.

Potential $V=\lambda^4 [ 1-({\phi \over M})^p ]$ ($p \geq 2$). For
$p=2$, the number of e-folds at $\phi_N$ before the end of
inflation is given by $N = \int^{t_{end}}_t H dt \simeq {1 \over
M^2_p} \int^{\phi_N}_{\phi_{end}} {V \over V'} d \phi \simeq (M^2
/ 2 M_p^2) \ln (M / \phi_N)$, or $\phi_N \simeq M exp ( -2 N M^2_p
/ M^2)$, where we put $\phi_{end} \simeq M$. The slow-roll
parameters are $\epsilon=2 M^2_p \phi_N^2 / M^4 = 2 (M_p^2 / M^2)
exp(-4 N (M_p^2 / M^2))$, $\eta=-2 (M_p^2 / M^2)$ and $\xi^2=0$.
In this case, $- \eta <<1$ implies $M_p << M$. For instance
assuming $M=10 M_p$ and $N=50$, we have $\epsilon = 0.0027$, $\eta
= - 0.02$, $n_s=0.96$ and $dn_s/ d \ln k = -0.001$. Because
$\epsilon$ is small, $\eta <0$ and $-32 \epsilon \eta \mu=0.0017
\mu > 0$, The spacetime noncommutative effects can not improve
this model. The same is true for $p>2$.

\noindent Case 4.

We consider the P-term inflation \lk. In unit with $M_p =1$, the
potential can be expressed as \eqn\pterm{V={g^2 \chi^2 \over 2}
\left(1+ {g^2 \over 8 \pi^2} ln {s^2 \over \chi}+ {f \over 8}
s^4+...\right), } here $0 \leq f \leq 1$ and special case $f=1$
corresponds to F-term inflation, $f=0$ corresponds to D-term
inflation. In P-term model the parameter $f$ is the contribution
from supergravity which permits a controllable running of the
scalar spectral index (no running in D-term case ). One should
note that $g^2$ in this model is not necessarily related to the
gauge coupling constant in GUTs. We have $g \geq 2 \times 10^{-3}$
and $\chi \simeq 10^{-5}$ \lk\ (constrained from CMB data), which
is a natural assumption (if $g << 2 \times 10^{-3}$, one has
exactly flat spectrum of density perturbations $n_s=1$. So we do
not consider this case.). Now we check whether this model can fit
the experimental data in noncommutative spacetime. For $g \geq 2
\times 10^{-3}$, the inflation is driven by the first term in
\pterm\ and we have \eqn\eppm{\epsilon=\half \left({g^2 \over 4
\pi^2 s}+{fs^3 \over2}\right)^2,  } \eqn\etpm{\eta=-{g^2 \over 4
\pi^2 s^2}+{3 f s^2 \over 2}, } \eqn\xipm{\xi^2=\left({g^2 \over 2
\pi^2 s^3}+3fs\right) \left({g^2 \over 4 \pi^2 s}+ {f s^3 \over
2}\right), } and $s^2_N = {g^2 N \over 2 \pi^2}$ where N is the
number of e-folds. According to \lk, for $g > 0.15$, inflation in
this model is too short, whereas for $g < 0.06$ one can ignore the
supergravity term $f s^4 / 8$ for the description of the last 60
e-folds of inflation. We expect that the spectral index $n$ runs
from $n < 1$ at small wavelengths to $n > 1$ at large wavelengths
in the intermediate regime $0.06 < g <0.15$. For example, we show
the spectral index and its running in the P-term inflation model
in figs. 5 and 6, where we choose $f=0.8$ and $g=0.08$.

\bigskip
{\vbox{{\epsfxsize=8cm
        \nobreak
    \centerline{\epsfbox{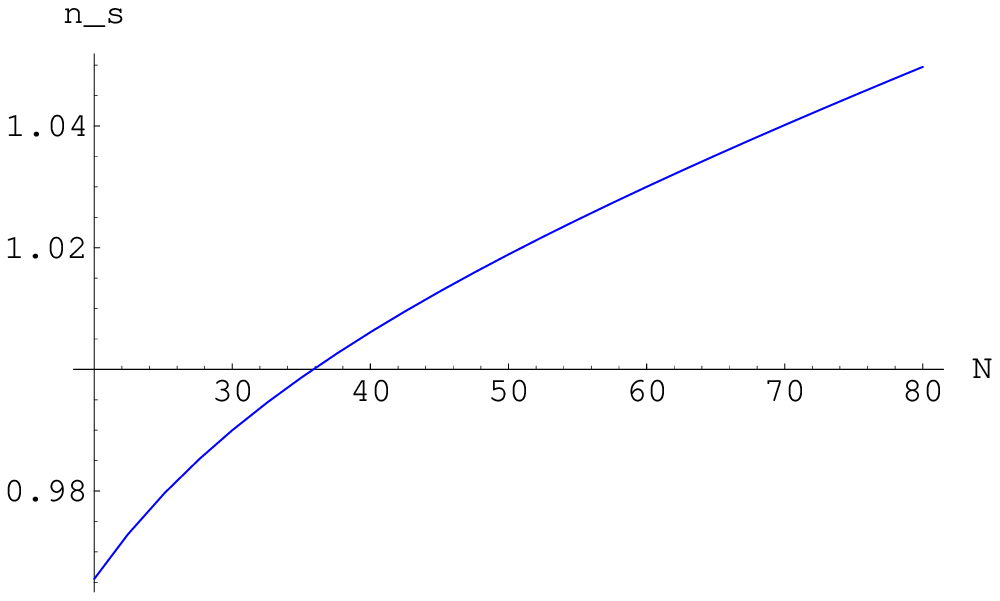}}
        \nobreak\bigskip
    {\raggedright\it \vbox{
{\bf Figure 5.}
{\it where $n_s$ in the spectral index and N is the number of
e-folds.}
 }}}}
    \bigskip}

\bigskip
{\vbox{{\epsfxsize=8cm
        \nobreak
    \centerline{\epsfbox{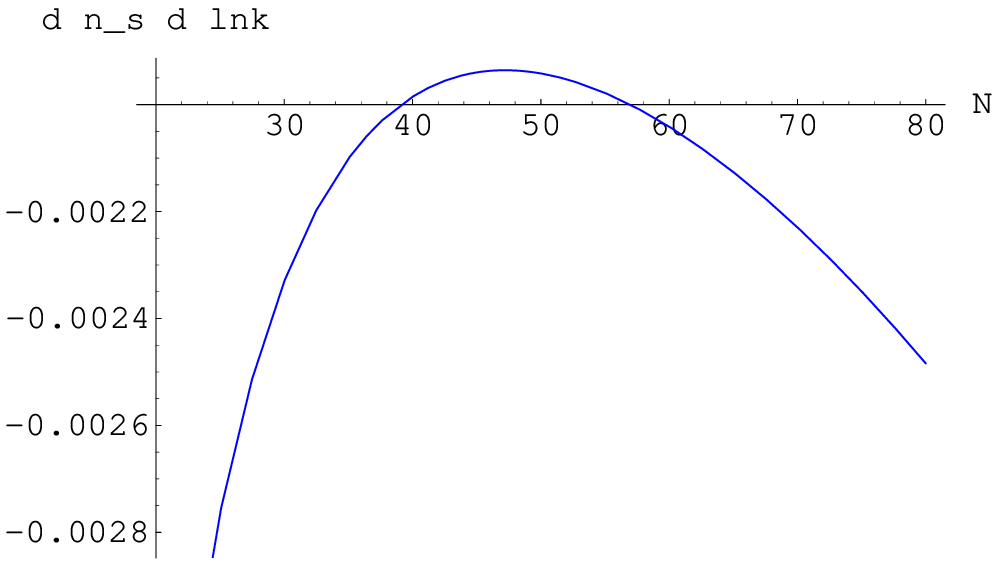}}
        \nobreak\bigskip
    {\raggedright\it \vbox{
{\bf Figure 6.}
{\it where $dn_s / dlnk$ in the running of the spectral index
and N is the number of e-folds.}
 }}}}
    \bigskip}
From fig. 5 and 6, we see that P-term inflation can not provide a
large enough running of the spectral index to fit the experiment
data. A similar result is obtained in \mmj. And also since
$\epsilon=2 \times 10^{-6} <<1 $, $\eta =0.0036$ for $N=40$ and
then $32 \epsilon \eta \mu = 7.2 \times 10^{-9} \mu$ which can be
ignored, we cannot expect the noncommutative effects improve this
model to fit the experiment data also.

To summarize, if we introduce the spacetime uncertainty relations into
cosmology, the classical evolution of the inflaton and the
universe is not different from the commutative case, since the
classical inflaton is homogeneous and the universe is spatially
isotropic and homogeneous. However this uncertainty relation
leads to the nonlocal coupling in time between the background and
the fluctuations. The time when the fluctuation mode crosses the
Hubble horizon is delayed for a smaller Hubble
constant. Thus the fluctuations will be smaller than those in
the absence of the uncertainty relation. This suppression implies
that the spectral index is larger than the one in the commutative case
for a long wave-length. With the
expansion of the universe, the cosmic scale becomes larger and
larger and the effects of the uncertainty relation become smaller
and smaller. So the fluctuations of the ultraviolet modes are not
different from the prediction of the standard theory.

The observations of WMAP brings about some more radical
suggestions, namely a running spectral index of the scalar
perturbation, making a transition from $n > 1$ on large scales to
$n < 1$ on small scales, and anomalously low quadrupole and
octupole. These two results are not anticipated in the usual
inflation models. In the last few months, many authors have
extensively discussed these results in \refs{\hlf-\hle, \shiu,
\mmj-\lowt}. The results of the cosmological observations are
still awaiting for further confirmation. In particular, the future
precise observations, for example measurements of $C^{TE}_l$ \dhl,
may be needed to show whether the anomalously low quadrupole and
octupole can be trusted or not. We also expect that the future
more precise measurements will provide a testing ground for
whether spacetime uncertainty is a viable physical model, and for
other new physics in the trans-Planckian regime.

\bigskip

Acknowledgments.
This work was supported by a
``Hundred People Project'' grant of Academia Sinica
and an outstanding young investigator award of NSF of China.

\listrefs
\end